\documentclass[12pt,aps,prd,superscriptaddress,showpacs,longbibliography,floatfix,nofootinbib]{revtex4-1}

\usepackage[utf8]{inputenc}
\usepackage{slashed}
\pdfoutput=1

\usepackage{color}
\usepackage{graphicx}   
\usepackage{bm}
\usepackage{amsmath}
\usepackage{amsfonts}
\usepackage{eufrak}
\usepackage{hyperref}

\newcommand{\be}{\begin{equation}}
\newcommand{\ee}{\end{equation}}
\newcommand{\ba}{\begin{eqnarray}}
\newcommand{\ea}{\end{eqnarray}}

\begin{document}

\title{Determining all thermodynamic transport coefficients for an interacting large N quantum field theory}

\author{Max Weiner}\affiliation{Department of Physics, University of Colorado, Boulder, Colorado 80309, USA}

\author{Paul Romatschke}
\affiliation{Department of Physics, University of Colorado, Boulder, Colorado 80309, USA}
\affiliation{Center for Theory of Quantum Matter, University of Colorado, Boulder, Colorado 80309, USA}

\begin{abstract}
  Thermodynamic transport coefficients can be calculated directly from quantum field theory without requiring analytic continuation to real time.
We determine all second-order thermodynamic transport coefficients for the uncharged N-component massless (critical) scalar field theory with quartic interaction in the large N limit, for any value of the coupling. We find that in the large N limit, all thermodynamic transport coefficients for the interacting theory can be expressed analytically in terms of the in-medium mass and sums over modified Bessel functions. We expect our technique to allow a similar determination of all thermodynamic transport coefficients for all theories that are solvable in the large N limit, including certain gauge theories.
\end{abstract}

\maketitle

\section{Introduction}

Transport coefficients govern the relaxation of a system back to equilibrium. Well-known examples of transport coefficients are conductivities, viscosities and diffusion constants \cite{landau2013fluid}.

In the context of relativistic quantum field theories, calculating the shear viscosity of QCD has been a long-standing goal, because relativistic nuclear collision experiments point to an unusually low value for this transport coefficient \cite{Florkowski:2017olj,Romatschke:2017ejr,Busza:2018rrf,Nagle:2018nvi}. However, calculating transport coefficients for any quantum field theory is hard. The QCD shear viscosity has been calculated for weak coupling using perturbation theory \cite{Arnold:2003zc,Ghiglieri:2018dib}, but extrapolating these calculations to coupling values required to make contact with experiment is wrought with large uncertainties. Monte Carlo simulations offer constraints on the shear viscosity value \cite{Meyer:2007ic}, but are plagued by systematic errors arising from the analytic continuation from Euclidean to Minkowski spacetime \cite{Burnier:2011jq}.

For quantum field theories with many components, large N expansions sometimes allow the calculation of exact transport properties directly from the field theory. For instance, for ${\cal N}=4$ super Yang-Mills theory, exact results are known in the infinite coupling limit using the AdS/CFT conjecture \cite{Policastro:2001yc}. For N-component scalar quantum field theory the shear viscosity has been calculated for \textit{any} interaction strength in both two and three spatial dimensions \cite{Aarts:2004sd,Romatschke:2021imm}.

There is, however, one class of transport coefficients that is considerably easier to calculate than others. These so-called ``thermodynamic'' transport coefficients possess the unique feature that analytic continuation from Euclidean to Minkowski space-time is not required for their determination. The moniker ``thermodynamic'' is a reminder of this property, which implies that thermodynamic transport coefficients can be obtained by calculating certain thermodynamic susceptibilities in equilibrium. While this feature would seem to suggest that thermodynamic transport coefficients are irrelevant for capturing the real-time evolution of a system, this is not the case. Instead, thermodynamic transport coefficients appear to take on a \textit{dual role}\footnote{This dual role is similar to the Einstein relation between diffusion constant and conductivity.} as controlling the real-time evolution to \textit{second} (or higher) order in derivatives\footnote{For novel insights into the relation between hydrodynamics and the gradient expansion, see e.g. \cite{Heller:2015dha,Romatschke:2017vte,Soloviev:2021lhs,Du:2021fok,Heller:2021yjh}.}, whereas the more well-known viscosities would control the first-order behavior. For this reason, this special class of transport coefficients is sometimes called second-order thermodynamic transport coefficients.

Kubo-like formulas for several thermodynamic transport coefficients have been found in the past two decades \cite{Baier:2007ix,Moore:2012tc,Buzzegoli:2017cqy}, but in a landmark paper Kovtun and Shukla \cite{Kovtun:2018dvd} showed how to obtain \textit{all} second-order thermodynamic transport coefficients from two-point correlation functions of the energy-momentum tensor. For relativistic uncharged fluids, this implies that all second-order transport coefficients can be obtained from knowledge of just three thermodynamic susceptibilities, which are called $f_1,f_2,f_3$ in Ref.~\cite{Kovtun:2018dvd}. The conditions relating $f_1-f_3$ and the thermodynamic transport coefficients generalizes previous results by independent methods \cite{Romatschke:2009kr,Bhattacharyya:2012nq,Jensen:2012jh}, which themselves have given rise to interesting physical interpretations \cite{Banerjee:2012iz,Banerjee:2012cr}.

Armed with the proper Kubo relations, all second-order thermodynamic transport coefficients are known for free uncharged relativistic scalar, vector and fermionic quantum field theories \cite{Moore:2012tc,Kovtun:2018dvd}, generalizing earlier results for individual thermodynamic transport coefficients, most notably $\kappa=-2 f_1$ \cite{Romatschke:2009ng}. (For free Dirac fermions, results have also been reported at finite density in Ref.~\cite{Shukla:2019shf}.)

For \textit{interacting} quantum field theories, determination of thermodynamic transport coefficients is still very hard, even though constraints for $\kappa$ have been reported for SU(3) Yang-Mills theory using Monte-Carlo simulations \cite{Philipsen:2013nea}.

Adding the large $N$ approximation to the box of tools allows for a much easier determination of thermodynamic transport coefficients for several classes of interacting quantum field theories. In particular, results have been calculated for large $N$, ${\cal N}=4$ super Yang-Mills theory at infinite coupling for both zero and finite density \cite{Bhattacharyya:2007vjd,Baier:2007ix,Finazzo:2014cna,Grozdanov:2014kva,Grieninger:2021rxd} using the AdS/CFT conjecture. At \textit{intermediate} coupling (not close to a free or infinitely strong coupled field theory), analytic results in the large $N$ limit have been found for $\kappa$ for the $O(N)$ model in \cite{Romatschke:2019gck} and for cold non-relativistic fermions at finite density in \cite{Lawrence:2022vwa}.

However, to date no complete determination of all second-order thermodynamic transport coefficients exist for an interacting quantum field theory at intermediate coupling\footnote{Note, however, the remarkable, yet unpublished, results by S.~Mahabir on this topic \cite{mahabir}}. This work is meant to fill this gap in the literature by calculating these transport coefficients for the $O(N)$ model.

\section{Calculation}

The theory we consider is defined by the curved-space action
\be
\label{action}
S=-\frac{1}{2}\int d^4x \sqrt{-g} \left[g^{\mu\nu} \partial_\mu \vec{\phi} \partial_\nu \vec{\phi} + \xi R \vec{\phi}^2+\frac{2\lambda}{N}\left(\vec{\phi}^2\right)^2\right]\,,
\ee
where $g_{\mu\nu}$ is the metric field in the mostly plus sign convention, $g\equiv {\rm det}g_{\mu\nu}$, \hbox{$\vec{\phi}=\left(\phi_1,\phi_2,\ldots,\phi_N\right)$} is an N-component scalar field, $\lambda$ is the coupling parameter and $\xi$ is a number that takes the value $\xi=\frac{1}{6}$ for the conformally coupled scalar. Energy-momentum tensor correlators in Minkowski space are calculated by twice differentiating $S$ with respect to $g_{\mu\nu}$ and subsequently setting the metric to the Minkowski metric. Without loss of generality, one may take metric perturbations to depend only on $g_{\mu\nu}(t,z)$, so that $x,y$ parametrize the transverse plane. With this choice, the calculation has been performed in Ref.~\cite{Kovtun:2018dvd}, finding the three thermodynamic susceptibilities given by
\ba
\label{f123}
f_1&=&-\frac{1}{2}\lim_{\bf k\rightarrow 0} \frac{\partial^2}{\partial k_z^2} G^{xy,xy}\,,\nonumber\\
f_2&=&\frac{1}{4}\lim_{\bf k\rightarrow 0} \frac{\partial^2}{\partial k_z^2} \left(G^{tt,tt}+2 G^{tt,xx}-4 G^{xy,xy}\right)\,,\nonumber\\
f_3&=&\frac{1}{4}\lim_{\bf k\rightarrow 0} \frac{\partial^2}{\partial k_z^2}\left(G^{tx,tx}+G^{xy,xy}\right)\,,
\ea
where for the action (\ref{action})
\be
G^{\mu\nu,\alpha\beta}(K)=\langle T^{\mu\nu} T^{\alpha\beta}\rangle_E(K)-\xi P^{\mu\nu,\alpha\beta,\rho\sigma} K_\rho K_\sigma \langle \vec{\phi}^2\rangle_E\,,
\ee
and $\langle \cdot \rangle_E$ denotes a Euclidean correlation function.
Here $K^\mu=\left(0,0,0,k_z\right)$ is the Minkowski 4-momentum evaluated at vanishing frequency and longitudinal momentum and  
\ba
P^{\mu\nu,\alpha\beta,\rho\sigma}&=&\eta^{\mu (\alpha} \eta^{\beta)(\sigma}\eta^{\rho)\nu}+\eta^{\mu(\rho}\eta^{\sigma)(\beta}\eta^{\alpha)\nu}-\eta^{\mu(\alpha}\eta^{\beta)\nu}\eta^{\rho \sigma}\nonumber\\
&&
-\eta^{\mu(\rho}\eta^{\sigma)\nu}\eta^{\alpha\beta}-\eta^{\mu\nu}\eta^{\alpha(\rho}\eta^{\sigma)\beta}+\eta^{\mu\nu}\eta^{\rho\sigma}\eta^{\alpha\beta}\,,
\ea
with $\eta^{\mu\nu}={\rm diag}(-,+,+,+)$ the Minkowski metric tensor and parentheses denote symmetrization.

Specifically,the required correlators then become
\ba
\label{line1}
G^{xy,xy}&=&\langle T^{xy} T^{xy}\rangle_E(K)+\frac{\xi}{2} k_z^2 \langle \vec\phi^2\rangle_E\,,\nonumber\\
G^{tx,tx}&=&\langle T^{tx} T^{tx}\rangle_E(K)-\frac{\xi}{2} k_z^2 \langle \vec\phi^2\rangle_E\,,\nonumber\\
G^{tt,xx}&=&\langle T^{tt} T^{xx}\rangle_E(K)+\xi k_z^2 \langle \vec\phi^2\rangle_E\,,\nonumber\\
G^{tt,tt}&=&\langle T^{tt} T^{tt}\rangle_E(K)\,.\nonumber\\
\ea

Therefore, in order to calculate the susceptibilities $f_1-f_3$, we need to calculate Euclidean correlators for the interacting field theory defined in (\ref{action}) in the large N limit.

\subsection{Calculating Euclidean Correlators}

The Euclidean version of the O(N) model is defined via the partition function $Z=\int {\cal D}\phi e^{-S_E}$ with the Euclidean action
\be
S_E=\int_0^\beta d\tau \int d^3x \left[\frac{1}{2}\partial_\mu \vec{\phi}\partial_\mu \vec{\phi}+\frac{\lambda}{N}\left(\vec{\phi}^2\right)^2\right]\,,
\ee
where $\beta=\frac{1}{T}$ is the inverse temperature of the system. We may perform a Hubbard-Stratonovic transformation by inserting $1=\int {\cal D}\sigma \delta(\sigma-\vec{\phi}^2)$ into the partition function and integrating out $\sigma$. This leads to
\be
Z=\int {\cal D}\phi {\cal D}\zeta e^{-S_{\rm eff}}\,,
\ee
with $\zeta$ an auxiliary field and
\be
\label{seff}
S_{\rm eff}=
\int_0^\beta d\tau \int d^3x \left[\frac{1}{2}\vec{\phi}\left(-\Box + i \zeta \right)\vec{\phi}
  +\frac{N}{16\lambda}\zeta^2\right]\,.
\ee
To capture the leading large N behavior for 0-point functions (e.g. the pressure), it is sufficient to use the R0 level resummation defined in Ref.~\cite{Romatschke:2019rjk}, which consists of splitting $\zeta$ into a zero mode and fluctuations and dropping the fluctuation contributions. Here we are interested in higher-point functions, so it is necessary to go to at least the R2 resummation level. At large N, this scheme is defined by adding and subtracting a self-energy for the fluctuations of $\zeta$, such that the action becomes $S_{\rm eff}=S_{2,0}+S_{2,I}$ with
\ba
\label{s2}
S_{2,0}&=&\frac{N \beta V \zeta_0^2}{16\lambda}+
\frac{1}{2}\int_X \vec{\phi}\left(-\Box + i \zeta_0 \right)\vec{\phi}
+ \frac{N}{2}\int_{X,Y}\zeta(X) \left[\frac{\delta(X-Y)}{8\lambda}+\Pi(X-Y)\right]\zeta(Y)\,,\nonumber\\
S_{2,I}&=&\frac{i}{2}\int_X \vec{\phi}^2\zeta-\frac{N}{2}\int_{X,Y}\zeta(X)\Pi(X-Y)
\zeta(Y)\,,
\ea
where $X\equiv (\tau,{\bf x})$, $\zeta$ is the fluctuation field without the zero mode fulfilling the constraint
\be
\label{cons}
\int_X \zeta(X)=0\,,
\ee
and $\Pi(X)$ is the self-energy that in the large N limit with the R2 level resummation is given by \cite{Romatschke:2019rjk}
\be
\Pi(X)=\frac{1}{2}G^2(X)\,,
\ee
with $G(X)=\langle \phi(X)\phi(0)\rangle_0$ the propagator for a single scalar field component (where $\langle \cdot \rangle_0$ indicates we are using the $S_{2,0}$ action). It is useful to also define the propagator for the auxiliary field, which is most easily done in Fourier space:
\be
\label{dd}
D(K)\equiv \langle \zeta \zeta\rangle_0=\frac{1-\frac{1}{\beta V}\delta(K)}{\frac{1}{8\lambda}+\Pi(K)}\,,
\ee
where $\beta V$ is the space-time volume of the system and the delta-function ensures that the zero mode constraint (\ref{cons}) is correctly taken into account.

Armed with this setup, it is straightforward to calculate Euclidean correlation functions. The easiest is
\be
\label{phi2}
\langle \vec{\phi}^2\rangle_E=N \left[G(0)-\langle \phi^2(X) S_{2,I} \rangle_0+\frac{1}{2}\langle \phi^2(X) S_{2,I}^2 \rangle_0+ \ldots\right]\,,
\ee
where as a reminder all Euclidean correlators are calculated using $S_{2,0}$. Since $S_{2,0}$ is quadratic in the fields $\phi$ and $\zeta$, Wick's theorem applies, and the first non-vanishing correction to $G(0)$ is $\frac{1}{2}\langle \phi^2(X) S_{2,I}^2 \rangle_0$. However, one can verify that this contribution vanishes because of the zero-mode constraint implied in (\ref{dd}). Similarly, all higher-order contributions in powers of $S_{2,I}$ that contribute to leading order in large N also vanish. As a consequence
\be
\langle \vec{\phi}^2\rangle_E=N G(0)+{\cal O}(N^0)\,,
\ee
where
\be
G(0)=T \sum_n \int \frac{d^3 {\bf k}}{(2 \pi)^3} \frac{1}{\omega_n^2+{\bf k}^2+i \zeta_0}\,, 
\ee
and $\omega_n=2 \pi n T$ are the bosonic Matsubara frequencies. With hindsight, writing the constant zero mode as
\be
i \zeta_0 \equiv m^2\,,
\ee
all the sum-integrals in $G(0)$ can be calculated analytically, finding  in dimensional regularization for $d=4-2\varepsilon$ dimensions \cite{Romatschke:2019gck}
\be
\label{g0}
G(0)=I(m)=-\frac{m^2}{16\pi^2 \varepsilon}-\frac{m^2}{16\pi^2}\ln \frac{\bar\mu^2e^1}{m^2} +  \frac{m}{2 \pi^2 \beta}\sum_{n=1}^\infty \frac{K_{1}(\beta n m)}{n}\,,
\ee
where $K_{\alpha}(x)$ denotes the modified Bessel function and $\bar\mu$ denotes the $\overline{\rm MS}$ renormalization scale.

It is worth pointing out that at this point,  the limit $\varepsilon\rightarrow 0$ implies $\langle \vec{\phi}^2\rangle_E$ is divergent on its own. Unless this divergence is exactly canceled by a similar divergence in the energy-momentum tensor correlators of opposite sign, the susceptibilities (\ref{f123}) would be ill-defined.

Furthermore, note that $m$ in (\ref{g0}) is not a free parameter, but is determined from the integration over the zero mode $\zeta_0$ in the path integral. Specifically, in the large N limit
\be
\label{zl}
Z=\int d\zeta_0 e^{N \beta V P(m)}\,,
\ee
where $P(m)=\frac{m^4}{16\lambda}-J(m)$ with $I(m)=2 \frac{d J(m)}{d m^2}$ so that \cite{Romatschke:2019gck}
\be
P(m)=\frac{m^4}{16\lambda}+\frac{m^4}{64\pi^2 \varepsilon}+\frac{m^4}{64\pi^2}\ln \frac{\bar\mu^2 e^{\frac{3}{2}}}{m^2}+\frac{m^2}{2 \pi^2 \beta^2}\sum_{n=1}^\infty \frac{K_2 (\beta n m)}{n^2}\,,
\ee
where here again $m=\sqrt{i \zeta_0}$.  Before doing the remaining integral over $\zeta_0$, the divergent contribution for $\varepsilon\rightarrow 0$ in $P(m)$ needs to be dealt with. This can be done via non-perturbative renormalization, introducing the renormalized coupling constant $\lambda_R$ as
\be
\label{ren}
\frac{1}{\lambda_R}=\frac{1}{\lambda}+\frac{1}{4\pi^2\varepsilon}\,.
\ee
so that
\be
P(m)=\frac{m^4}{16 \lambda_R}+\frac{m^4}{64\pi^2}\ln \frac{\bar\mu^2 e^{\frac{3}{2}}}{m^2}+\frac{m^2}{2 \pi^2 \beta^2}\sum_{n=1}^\infty \frac{K_2 (\beta n m)}{n^2}\,.
\ee
In the large N limit, the remaining integral over $\zeta_0$ in (\ref{zl}) can be done exactly using the saddle point method, leading to
\be
Z=e^{N \beta V P(m)}\,,
\ee
with $m$ now determined by the condition $0=P^\prime(m)$.
Integrating up the renormalization condition (\ref{ren}) leads to the running coupling
\be
\frac{1}{\lambda_R(\bar\mu)}=\frac{1}{4\pi^2}\ln \frac{\Lambda_{LP}^2}{\bar\mu^2}\,,
\ee
where $\bar\mu=\Lambda_{LP}$ is the location of the Landau pole of the theory. The saddle point condition thus can be written as
\be
\label{gap}
0=P^\prime(m)=\frac{m^3}{16\pi^2}\ln \frac{\Lambda_{LP}^2 e^{1}}{m^2}-\frac{m^2}{2 \pi^2 \beta} \sum_{n=1}^\infty \frac{K_1\left(\beta n m\right)}{n}\,.
\ee

At zero temperature, this gap equation has two solutions, $m=0$ and $m=\Lambda_{LP} \sqrt{e}$. Since the second solution involves energy scales above the Landau pole, it can safely be discarded as unphysical, so that only $m=0$ remains at zero temperature.

\begin{figure}
  \includegraphics[width=.7\linewidth]{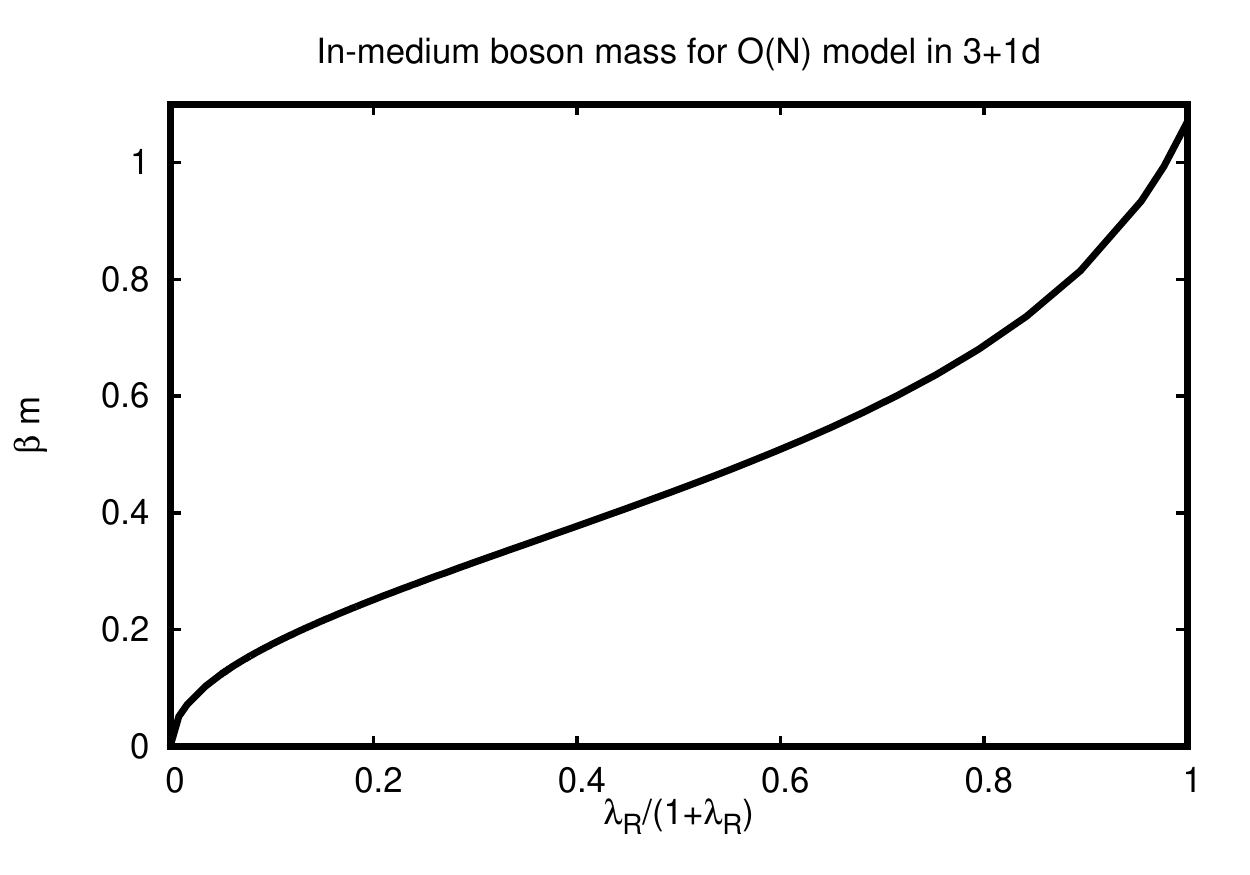}
  \caption{Non-trivial solution to the gap equation (\ref{gap}) for the in-medium mass $m$  as a function of the renormalized coupling \label{fig1} with the choice $\bar\mu=2 \pi T$. Note that the x-axis is compactified so as to show the entire interval $\lambda_R\in[0,\infty)$. }
  \end{figure}

For general temperature $T\ll \Lambda_{LP}$, this gap equation possesses three solutions. One is the trivial solution $m=0$, and another one is a solution close to $m \simeq \Lambda_{LP} \sqrt{e}$, which we again discard as unphysical. Finally, there is a non-trivial solution $m\neq 0$ that vanishes at zero temperature but is non-vanishing at finite temperature (see Fig.~\ref{fig1}). The temperature dependence of this solution $m$ encodes interactions of the theory in the leading large N limit. All correlation functions (such as Eq.~(\ref{g0})) are evaluated at this particular mass scale. Finally, note that the saddle-point equation obeyed by this non-trivial solution can be written as
\be
\label{non-trivial-gap}
m^2=4 \lambda I(m)\,,
\ee
in accordance with earlier findings \cite{Romatschke:2019gck}.

The pressure evaluated at the solution of the gap equation may be written as
\be
P(m)=\frac{N m^4}{128 \pi^2}\left(1+16 \sum_{n=1}^\infty \frac{K_3(\beta n m)}{\beta n m}\right)\,,
\ee
from which the energy density follows as
\be
\epsilon(m)=3 P(m) -\frac{N m^4}{32 \pi^2}\,.
\ee
After a little algebra, we find the speed of sound squared $c_s^2\equiv \frac{dP}{d\epsilon}$ to be given by
\be
\label{cs2}
c_s^2=\left(3+\frac{\sum_{n=1}^\infty K_2\left(\beta n m\right)}{\left(1-4\sum_{n=1}^\infty K_2\left(\beta n m\right)\right)\sum_{n=1}^\infty \frac{K_3(\beta n m)}{\beta n m}}\right)^{-1}\,.
\ee
As discussed in detail in Ref.~\cite{Romatschke:2019gck}, one has $c_s^2\geq \frac{1}{3}$ for all temperatures.

\subsection{Calculating Euclidean Off-Diagonal Energy-Momentum Tensor Correlators}

Besides $\langle \vec{\phi}^2\rangle_E$, the other ingredient necessary to calculate $f_1,f_2,f_3$ are the energy-momentum tensor correlators. The energy momentum tensor in Minkowski space for the action (\ref{action})  is given by
\be
\label{emt}
T^{\mu\nu}=\partial^\mu \vec{\phi}\partial^\nu \vec{\phi}-\eta^{\mu\nu} \left[\frac{1}{2}\partial_\mu \vec{\phi}\partial^\mu \vec{\phi}+\frac{\lambda}{N}\left(\vec{\phi}^2\right)^2\right]\,.
\ee
Clearly, the off-diagonal components $\mu\neq \nu$ of the energy-momentum tensor are simpler, and we start calculating two-point correlation functions for those first.

Specifically, upon analytically continuing from real to imaginary time $t\rightarrow -i \tau$, we have
\ba
\langle T^{xy}(X) T^{xy}(0)\rangle_E&=&\langle \partial_x \phi_i(X) \partial_y \phi_i(X)\partial_x \phi_j(0)  \partial_y \phi_j(0)\rangle_E\,,\nonumber\\
\langle T^{tx}(X) T^{tx}(0)\rangle_E&=&-\langle \partial_\tau \phi_i(X) \partial_x \phi_i(X)\partial_\tau \phi_j(0)  \partial_x \phi_j(0)\rangle_E\,.
\ea
Expanding the exponential of the effective action (\ref{seff}) in powers of $S_{2,I}$, one finds that for the connected off-diagonal correlators, only the one-loop bubble in the large N limit is non-vanishing. In Fourier space, then
\ba
\langle T^{xy} T^{xy}\rangle_E (K)&=&2 N T \sum_n \int \frac{d^3p}{(2\pi)^3} {\bf p}_x^2 {\bf p}_y^2 G(P)G(P+K) + {\cal O}(N^0)\,,\nonumber\\
\langle T^{tx} T^{tx}\rangle_E(K)&=&-2 N T \sum_n \int \frac{d^3p}{(2\pi)^3} {\bf p}_x^2 \omega_n^2 G(P)G(P+K)+ {\cal O}(N^0)\,,
\ea
where again $K^\mu=\left(0,0,0,k_z\right)$ for our purposes and
\be
G(P)=\frac{1}{\omega_n^2+{\bf p^2}+m^2}\,,
\ee
is the propagator in momentum space with the interaction-dependent $m$ given by the solution of (\ref{gap}). Since we are interested in the ${\cal O}(k_z^2)$ piece of the correlator, we can expand the propagators in the loop as
\be
G(P)G(P+K)={\cal O}(k_z^0)+{\cal O}(k_z^1)+k_z^2 \left[-G^3(P)+4 p_z^2 G^4(P)\right]+{\cal O}(k_z^3)\,.
\ee
Performing the angular averages and repeatedly using $G^2(P)=-\partial_{m^2} G(P)$ to rewrite the powers of the propagator, we find
\ba
\partial_{k_z}^2\langle T^{xy} T^{xy}\rangle_E (0)&=&4 N T \sum_n \int \frac{d^3p}{(2\pi)^3} \left[-\frac{p^4}{30}\partial^2_{m^2}-\frac{2 p^6}{315} \partial^3_{m^2}\right]G(P) \,,\nonumber\\
\partial_{k_z}^2\langle T^{tx} T^{tx}\rangle_E(0)&=&-4 N T \sum_n \int \frac{d^3p}{(2\pi)^3} \omega_n^2 \left[-\frac{p^2}{6}\partial^2_{m^2} -\frac{2 p^4}{45} \partial^3_{m^2}\right]G(P)\,.
\ea
Evaluating the standard thermal sums then gives
\ba
\partial_{k_z}^2\langle T^{xy} T^{xy}\rangle_E (0)&=&2 N \int \frac{d^3p}{(2\pi)^3} \left[-\frac{p^4}{30}\partial^2_{m^2}-\frac{2 p^6}{315} \partial^3_{m^2}\right]\frac{1}{E_p}\left(1+2 n_B(\beta E_p)\right)\,,\nonumber\\
\partial_{k_z}^2\langle T^{tx} T^{tx}\rangle_E(0)&=&2 N \int \frac{d^3p}{(2\pi)^3} \left[-\frac{p^2}{6}\partial^2_{m^2} -\frac{2 p^4}{45} \partial^3_{m^2}\right]E_p\left(1+2 n_B(\beta E_p)\right)\,,
\ea
where $E_p=\sqrt{p^2+m^2}$ and $n_B(x)=\frac{1}{e^x-1}$. The terms independent from $n_B$ are UV divergent and are treated using dimensional regularization in $\overline{\rm MS}$. We find
\ba
\partial_{k_z}^2\langle T^{xy} T^{xy}\rangle_E (0)&=&\frac{2}{3} \frac{Nm^2}{64\pi^2} \left( \frac{1}{\epsilon} + \frac{53}{105} +  \ln \left( \tfrac{\bar{\mu}^2}{m^2} \right) - 8 \sum_{n=1}^{\infty} \frac{K_1(n\beta m)}{n\beta m} \right)\,,\\
\partial_{k_z}^2\langle T^{tx} T^{tx}\rangle_E(0)&=&-\frac{2}{3}\frac{Nm^2}{64\pi^2} \left( \frac{1}{\epsilon} + \frac{17}{15} + \ln \left( \tfrac{\bar{\mu}^2}{m^2} \right) + \sum_{n=1}^{\infty} \left\{ 8K_0(n\beta m) + \frac{8K_1(n\beta m)}{n\beta m} \right\} \right)\,,\nonumber
\ea
so that together with (\ref{phi2}), (\ref{g0}), (\ref{line1}) we have for $\xi=\frac{1}{6}$
\ba
\lim_{\bf k\rightarrow 0}\partial_{k_z}^2 G^{xy,xy}&=&-\frac{13Nm^2}{2520\pi^2} \,,\nonumber\\
\lim_{\bf k\rightarrow 0} \partial_{k_z}^2 G^{tx,tx}&=&-\frac{Nm^2}{96\pi^2} \left(  \frac{2}{15} + 8\sum_{n=1}^{\infty}  K_2\left(n\beta  m  \right)   \right) \,.
  \ea
  Note that all divergences, as well as all explicit dependencies on the renormalization scale $\bar\mu$, have canceled for the conformally coupled scalar. Also, we recall that for $\xi\neq \frac{1}{6}$, the divergences do not cancel, which suggests that coupling a scalar to gravity is only consistent in the conformal case, cf. Refs.~\cite{Romatschke:2019gck,Kuipers:2021jlh}.

  \subsection{Calculating Euclidean Diagonal Energy-Momentum Tensor Correlators}

  Besides the off-diagonal components of the energy-momentum tensor, we also need correlators of the diagonal components, such as
  \ba
  T^{tt}=\frac{1}{2}\left[-\partial_\tau \vec{\phi}\cdot \partial_\tau \vec{\phi}
    +\partial_x \vec{\phi}\cdot \partial_x \vec{\phi}+\partial_y \vec{\phi}\cdot \partial_y \vec{\phi}+\partial_z \vec{\phi}\cdot \partial_z \vec{\phi}\right]
+\frac{\lambda}{N}\left(\vec{\phi}^2\right)^2\,.
\ea
While it appears at first glance that the quartic term is subleading in the large $N$ limit, this turns out to be incorrect since loops involving $N$ components of $\phi$ can compensate the $\frac{1}{N}$ term. Using shorthand notation
\be
\langle \partial_\tau \phi_i(X) \partial_\tau \phi_i(X)  \partial_\tau \phi_j(0) \partial_\tau \phi_j(0)\rangle= \langle \tau\tau\ \tau\tau\rangle\,,
\ee
to denote correlators involving derivatives, we get contributions such as
\ba
\label{baa1}
\langle\tau \tau \tau \tau\rangle_E  &=&  \langle \tau\tau \tau\tau\rangle_0+\frac{1}{2} \langle \tau \tau \tau \tau S_{2,I}^2\rangle_0+{\cal O}(N^0)\,,\nonumber\\
\frac{\lambda}{N}\langle\tau \tau (\vec{\phi}^2)^2\rangle_E  &=& \frac{\lambda}{N}\langle \tau\tau (\vec{\phi}^2)^2\rangle_0+\frac{\lambda}{2 N} \langle \tau \tau \tau \tau S_{2,I}^2 (\vec{\phi}^2)^2\rangle_0+{\cal O}(N^0)\,,\\
\frac{\lambda^2}{N^2} \langle (\vec{\phi}^2)^2 (\vec{\phi}^2)^2\rangle_E&=& \frac{\lambda^2}{N^2} \langle (\vec{\phi}^2)^2 (\vec{\phi}^2)^2\rangle_0 +\frac{\lambda^2}{2 N^2}\langle (\vec{\phi}^2)^2 (\vec{\phi}^2)^2 S_I^2\rangle_0+{\cal O}(N^0)\,.\nonumber
\ea
It can be checked that the polarization tensor counter-term in (\ref{s2}) cancels all higher power insertions of $S_{2,I}$ in the large N limit, so that (\ref{baa1}) is exact in the large N limit. The calculation of the correlation function proceeds as in the preceding subsection, and we find in Fourier space to leading order in large $N$
\ba
\langle \tau\tau \tau\tau\rangle_E(K) &=&2 N T \sum_n \int \frac{d^3 p}{(2\pi)^3}
  \omega_n^4 G(P)G(P+K)-2N D(K) B_\tau^2(K)\,,\nonumber\\
\frac{\lambda}{N}\langle \tau\tau (\vec{\phi}^2)^2\rangle_E(K)&=& N m^2 B_\tau(K)-2N m^2 D(K) \Pi(K) B_\tau(K)\,, \\
\frac{\lambda^2}{N^2} \langle (\vec{\phi}^2)^2 (\vec{\phi}^2)^2\rangle_E(K) &=&N m^4 \Pi(K)-2N m^4D(K) \Pi^2(K)\,, \nonumber
\ea
where we used the non-trivial gap equation (\ref{non-trivial-gap}) to rewrite the coupling-constant dependence in terms of the in-medium mass shown in Fig.~\ref{fig1}. Here and in the following, the short-hand notation introduced is
\ba
B_{\underline{\mu}}(K)\equiv T \sum_n \int \frac{d^3 p}{(2\pi)^3} P_{\underline{\mu}} \left(P_{\underline{\mu}}+K_{\underline{\mu}}\right)  G(P)G(P+K)\,,
\ea
where $\underline{\mu}$ indicates that there is no summation over the index, for instance $B_\tau(K)=T \sum_n \omega_n \left(\omega_n+k_0\right) G(P)G(P+K)$. Given these results, we can write the large $N$ limit of the diagonal energy-momentum tensor correlators as
\ba
\langle T^{tt}T^{tt}\rangle_E(K)&=&N T \sum_n \int \frac{d^3 p}{(2\pi)^3}G(P)G(P+K)\bigg(\frac{\omega_n^4}{2}+\frac{{\bf p}_x^4}{2}+\frac{{\bf p}_y^4}{2}+\frac{{\bf p}_z^2({\bf p}_z+{\bf k}_z)^2}{2} \nonumber\\
&&-\omega_n^2 {\bf p}_x^2-\omega_n^2 {\bf p}_y^2-\omega_n^2 {\bf p}_z({\bf p}_z+{\bf k}_z)+{\bf p}_x^2{\bf p}_y^2+{\bf p}_x^2{\bf p}_z ({\bf p}_z+{\bf k}_z)+{\bf p}_y^2{\bf p}_z ({\bf p}_z+{\bf k}_z)\bigg) \nonumber\\
&&-Nm^2B_{\tau}+Nm^2B_x+Nm^2B_y+Nm^2B_z+Nm^4\Pi(K) \\
&&-\frac{N}{2}D(K)\left(B_\tau-2B_x-B_z-2m^2\Pi(K)\right)^2 \nonumber\\
\langle T^{tt}T^{xx}\rangle_E(K)&=&N T \sum_n \int \frac{d^3 p}{(2\pi)^3}
G(P)G(P+K)\left(\frac{\omega_n^4}{2}-\omega_n^2 {\bf p}_x^2-{\bf p}_y^2 {\bf p}_z ({\bf p}_z+{\bf k}_z)-\frac{{\bf p}_z^2 ({\bf p}_z+{\bf k}_z)^2}{2}\right)\nonumber\\
&&-N m^2 B_y-N m^2 B_z-Nm^4 \Pi(K)\nonumber\\
&&+\frac{N}{2} D(K) \left(B_\tau+B_z+2 m^2 \Pi(K)\right)\left(B_z+2 B_x-B_\tau+2 m^2 \Pi(K)\right)\,.\nonumber
\ea
Performing the integrals and using $D(K=0)=0$ from (\ref{dd}), we find that all the divergences in the correlation functions (\ref{line1}) cancel up to ${\cal O}(k_z^2)$ for $\xi=\frac{1}{6}$ so that the susceptibilities $f_1,f_2,f_3$ are all finite. Specifically, we find
\ba
\label{line2}
\lim_{\bf k\rightarrow 0}\partial^2_{k_z}G^{tt,tt}&=&\frac{Nm^2}{32\pi^2}\left(\frac{2}{3} - \frac{8}{3}\sum_{n=1}^{\infty} \left\{ n\beta m K_1(n\beta m) + 4K_2(n\beta m) \right\} \right)\nonumber\\
&&-\frac{4Nm^2}{3\pi^2}\frac{1+2\sum_{n=1}^{\infty}n \beta m K_1(n\beta m)}{\left[1-4\sum_{n=1}^{\infty}K_2(n\beta m)\right]^2}\left(\sum_{n=1}^{\infty}K_2(n\beta m)\right)^2 ,\nonumber\\
\lim_{\bf k\rightarrow 0}\partial^2_{k_z}G^{tt,xx}&=&\frac{Nm^2}{32\pi^2}\left( -\frac{14}{45} + \frac{16}{3}\sum_{n=1}^{\infty}K_2(n\beta m) \right)\nonumber\\
&&-\frac{Nm^2}{9\pi^2} \frac{1+2\sum_{n=1}^{\infty}n \beta m K_1(n\beta m)}{\left[1-4\sum_{n=1}^{\infty}K_2(n\beta m)\right]^2} \sum_{n=1}^{\infty} K_2(n\beta m) \,,
\ea
and as a consequence (\ref{f123}) leads to
\ba
\label{mainresults}
f_1&=&\frac{13Nm^2}{5040\pi^2}\,, \nonumber\\
f_2&=&\frac{37Nm^2}{6720\pi^2} - \frac{Nm^2}{48\pi^2}\sum_{n=1}^{\infty}n \beta m K_1(n\beta m)\nonumber\\
&&-\frac{2Nm^2}{9\pi^2}\frac{1+2\sum_{n=1}^{\infty}n \beta m K_1(n\beta m)}{\left[1-4\sum_{n=1}^{\infty}K_2(n\beta m)\right]^2}\left( \sum_{n=1}^{\infty}K_2(n\beta m)+6\left(\sum_{n=1}^{\infty}K_2(n\beta m)\right)^2\right)\,,\nonumber\\
f_3&=&-\frac{11 N m^2}{6720 \pi ^2} - \frac{Nm^2}{48\pi^2} \sum_{n=1}^{\infty}  K_2\left(n \beta  m  \right)\,.
\ea
These are our main results.

\begin{figure}
  \includegraphics[width=.7\linewidth]{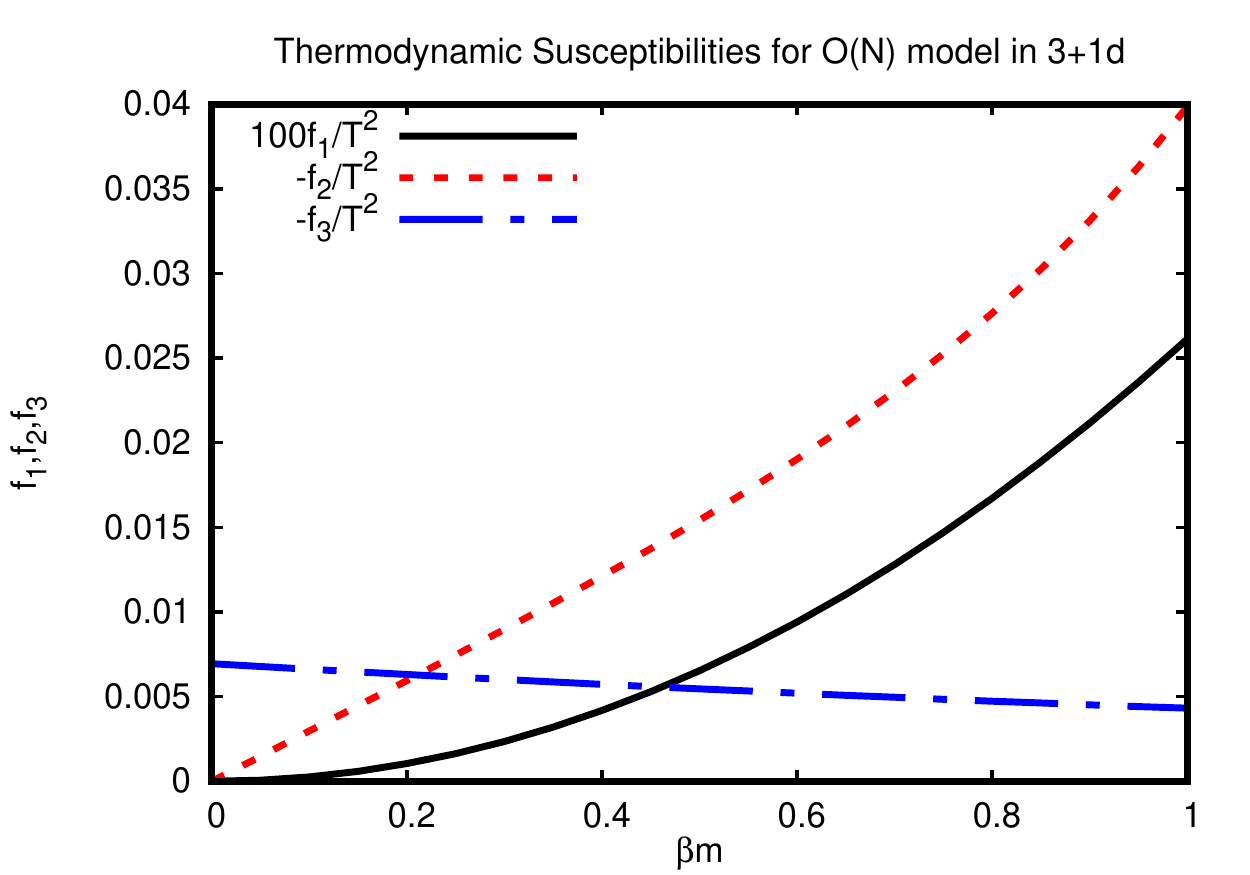}
  \caption{Thermodynamic susceptibilities $f_1,f_2,f_3$ as a function of in-medium boson mass (\ref{non-trivial-gap}). Results are independent from the choice of the renormalization scale $\bar\mu$.\label{fig2}}
  \end{figure}

Note that for the conformally coupled scalar, the susceptibilities $f_1,f_2,f_3$ are divergence-free, and that all explicit dependencies on the renormalization scale $\bar\mu$ have canceled. For weak coupling or small temperatures, the gap equation (\ref{non-trivial-gap}) implies $\beta m\ll 1$. In this limit, we find $f_1=f_2=0,f_3=-\frac{N T^2}{144}$, matching the results for $N$ free scalars found in Refs.~\cite{Moore:2012tc,Kovtun:2018dvd}. Our results (\ref{mainresults}) are well defined and can be numerically evaluated for all $T\ll \Lambda_{LP}$ and couplings (see Fig.~\ref{fig2}).

\section{Discussion}

In this work, we have performed an exact calculation of the thermodynamics susceptibilities for an interacting quantum field theory in the large N limit for any value of the interaction. Since the quantum field theory we considered possesses a Landau pole in the large N limit, our results only make physical sense in the effective field theory sense, e.g. for energy scales much below the Landau pole. Despite this limitation, one can expect our results to be physically meaningful for weak and intermediate interactions.

Knowledge of the three susceptibilities $f_1,f_2,f_3$ (\ref{mainresults}) as well as the speed of sound squared (\ref{cs2}) for an interacting theory allows us to extract all second order thermodynamic transport coefficients for an uncharged relativistic fluid. In particular, using the translations worked out in Ref.~\cite{Kovtun:2018dvd}, we are able to evaluate all 8 transport coefficients defined in Ref.~\cite{Romatschke:2017ejr} as
\ba
\label{all}
\kappa&=&-2 f_1\,,\nonumber\\
\kappa^*&=&f_1^\prime-2 f_1\,,\nonumber\\
\lambda_3&=&2 (f_1^\prime-4 f_3)\,,\nonumber\\
\lambda_4&=&c_s^4\left(4 f_1^\prime+2 f_1^{\prime\prime}-2 f_2\right)\,,\nonumber\\
\xi_3&=&-2 c_s^2 \left(f_1^\prime-f_2-3f_3+f_3^\prime\right)-\frac{2}{3}\left(f_3+2 f_1^\prime\right)\,,\nonumber\\
\xi_4&=&-c_s^6\left(4 f_1^\prime+2 f_1^{\prime\prime}-2 f_2^\prime-f_2\right)-\frac{1}{3}c_s^4\left(4f_1^{\prime \prime}+2 f_1^\prime-f_2\right)\,,\nonumber\\
\xi_5&=&c_s^2\left(f_1-f_1^\prime\right)+\frac{1}{3}f_1\,,\nonumber\\
\xi_6&=&-2 c_s^2 \left(f_2-f_1-f_1^\prime\right)+\frac{2}{3}\left(2 f_1^\prime-f_1\right)\,,
\ea
where the primes denote $f^\prime =T \frac{d f_1}{dT}$, $f^{\prime \prime}=T^2 \frac{d^2 f_1}{dT^2}$, etc. These in turn may be helpful in studying conjectured relations among transport coefficients such as \cite{Haack:2008xx,Kleinert:2016nav}
\be
\label{conj}
f_1=\kappa^*-\kappa\overset{?}{=}2 \lambda_1\,,
\ee
where $\lambda_1$ is a \textit{non-thermodynamic} (e.g. real-time) transport coefficient. As discussed in the introduction, real-time transport coefficients are generally hard to calculate for quantum field theories, so relations such as (\ref{conj}) could be shaped into powerful tools to study real-time transport for interacting quantum field theories.

\begin{figure}
  \includegraphics[width=.7\linewidth]{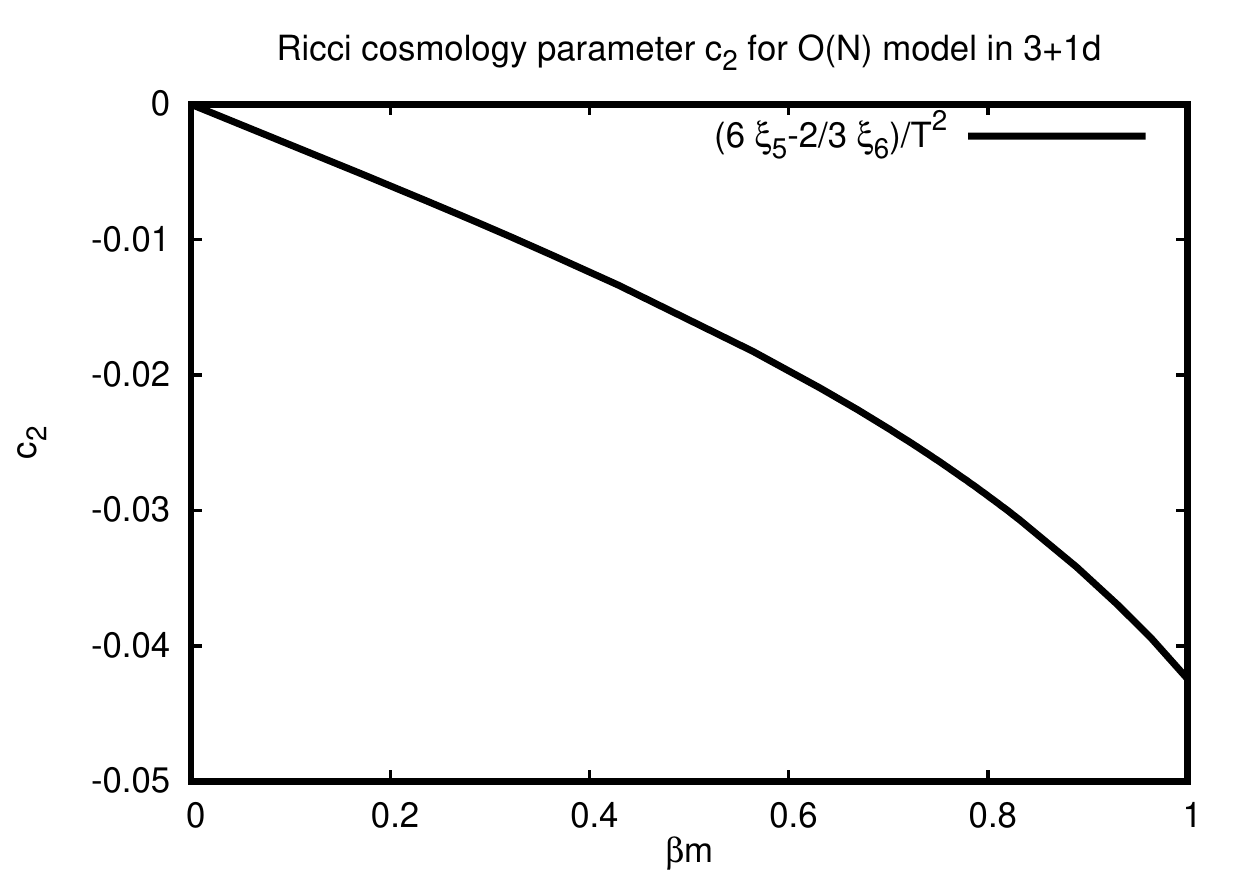}
  \caption{The combination (\ref{c2}) defined in the main text as a function of the in-medium mass $\beta m$. We find that the sign of $c_2$ is always negative for the interacting O(N) model, which would imply an inflationary period in cosmology without cosmological constant. \label{fig3}}
  \end{figure}

Another application of (\ref{all}) is direct access to the combination
\be
\label{c2}
c_2\propto \frac{1}{T^2}\left(6 \xi_5-\frac{2}{3} \xi_6\right)\,,
\ee
defined in Ref.~\cite{Baier:2019ciw}. It was found in this reference that if $c_2<0$, the universe would undergo an inflationary phase \textit{without the presence of a cosmological constant} (e.g. with just standard model of physics ingredients). We plot the combination $c_2$ in Fig.~\ref{fig3} and indeed find that $c_2<0$ for all coupling values. This could further strengthen the case of a Ricci-cosmology-type scenario for the early universe \cite{Caroli:2021mjg} or illuminate further the behavior of Einstein equations in matter in AdS spacetimes, as pointed out in Ref.~\cite{Kovtun:2019wjz}.

In addition to these two specific questions, we hope that the full knowledge of all thermodynamic transport coefficients presented in this work can serve as a treasure-trove for testing conjectures and putting constraints on models from a wide range of fields, in particular as researchers start to study more of the implications of second-order transport (cf. Ref.~\cite{Andersson:2020phh}).

Moreover, the fact that divergences in the susceptibilities cancel only in the case of the conformally coupled scalar may help also in the context of research in quantum gravity, cf. Ref.~\cite{Kuipers:2021jlh}.

Going forward, we expect that our results may be generalized to calculate thermodynamic transport coefficients for all quantum field theories that are 'solvable' in the large N limit, in particular also in other dimensions \cite{Grable:2022swa}, with fermions \cite{Pinto:2020nip}, supersymmetric theories \cite{DeWolfe:2019etx} and for certain tensor theories, cf. \cite{Klebanov:2018fzb}.

There is plenty of work left to do.

  \section{Acknowledgments}

  This work was supported by the Department of Energy, DOE award No DE-SC0017905. We would like to thank Ashish Shukla for useful discussions, and S. Mahabir for collaboration in the early stages of this project.

\bibliography{ufg}
\end{document}